\documentclass[aps,groupedaddress,amssymb,amsmath,amsbsy,amsfonts,twocolumn,pra]{revtex4-1}

\usepackage{amssymb,amsmath,amsthm,color,graphicx,times,graphicx}
\usepackage{hyperref}
\usepackage{enumerate}
\usepackage{subfigure}
\usepackage{dsfont}
\usepackage{times}
\usepackage{txfonts}
\usepackage{bbold}
\usepackage{color}

\newcommand{\affnot}{Centre for the Mathematics and Theoretical Physics of Quantum Non-Equilibrium Systems (CQNE),
School of Mathematical Sciences,  University of Nottingham, Nottingham NG7 2RD, United Kingdom}

\renewcommand{\Re}{\mathrm{Re}}
\renewcommand{\Im}{\mathrm{Im}}

\theoremstyle{plain}

\newtheorem{theorem}{Theorem}

\theoremstyle{definition}

\begin{document}

\title{Multiparameter Gaussian Quantum Metrology}
\author{Rosanna Nichols}\email{pmxrn1@nottingham.ac.uk}\affiliation{\affnot}
\author{Pietro Liuzzo-Scorpo}\email{pmxpl2@nottingham.ac.uk}\affiliation{\affnot}
\author{Paul A. Knott}\email{paul.knott@nottingham.ac.uk}\affiliation{\affnot}
\author{Gerardo Adesso}\email{gerardo.adesso@nottingham.ac.uk}\affiliation{\affnot}

\begin{abstract}
We investigate the ultimate precision achievable in Gaussian quantum metrology. We derive general analytical expressions for the quantum Fisher information matrix and for the measurement compatibility condition, ensuring asymptotic saturability of the quantum Cram\'er-Rao bound, for the estimation of multiple parameters encoded in multimode Gaussian states. We then apply our results to the joint estimation of a phase shift and two parameters characterizing Gaussian phase covariant noise in optical interferometry. In such a scheme, we show that two-mode displaced squeezed input probes with optimally tuned squeezing and displacement fulfil the measurement compatibility condition and enable the simultaneous estimation of all three parameters, with an advantage over individual estimation schemes that quickly rises with increasing mean energy of the probes. \end{abstract}
\date{\today}

\maketitle

\section{Introduction}

The exploitation of quantum effects for enhancements in sensing and precision measurements stands as one of the linchpins of the current {\it  quantum technology} revolution \cite{Caves1981,Huelga1997,Giovannetti2006,PARIS2009,Giovannetti2011,Dowling2015,Toth2014,DattaReview,Pezze2016,Braun2017}. The applications of quantum metrology range from fundamental science, such as improving time and frequency standards \cite{Giovannetti2004,Katori2011}, advancing the sensitivity of gravitational wave interferometry \cite{Schnabel2010,Schnabel2013}, and probing space-time parameters in quantum field theory \cite{Aspachs2010,Ahmadi2014},  to more applied scenarios, such as navigation \cite{Cai2013,Komar2014}, remote sensing \cite{Boto2000,Dowling2015}, thermometry \cite{Correa2015,DePasquale2016}, spectroscopy \cite{Schmitt2017,Boss2017}, super-resolution imaging \cite{Tsang2016,Nair2016,Lupo2016}, magnetic field detection for biomedical diagnostics \cite{Taylor2014,Bonato2015}, and plenty more to come.

In many of these settings, the problem can be modelled as the estimation of unknown parameters encoded on a probe field initialized in a continuous variable (CV) Gaussian state, i.e., a Gibbs ensemble of a quadratic Hamiltonian \cite{Braunstein2005,Weedbrook2012,Adesso2014,Serafini2017}. If the (unitary or noisy) channel imprinting the parameters preserves the Gaussianity of the input state, the setting is overall referred to as {\em Gaussian quantum metrology}.
Several works analyzed instances of Gaussian quantum metrology, including the estimation of single or multiple parameters using single-mode or multimode probes
\cite{Caves1981,Monras2006,Monras07,AdessoAnno09,Gaiba09,Bellomo2009,Aspachs2010,PhysRevA.87.012107,Monras2011,
Pinel12,Pinel13,Monras13,Gao2014,Jiang2014,Adesso2014,Ahmadi2014,Correa2015,Safranek2015,Friis2015,Banchi15,Rigovacca2015,Rafau2015,Safranek2016,Marian2016,PhysRevA.95.012305,Gagatsos2016,Liu2016,
Tsang2016,Nair2016,Lupo2016,Liu2016,Ragy2016,Pearce2017,Zhuang2017,Proctor2017,Pirandola2017,Giovannetti2011,DattaReview,Braun2017,Serafini2017}. However, to the best of our knowledge, no general method is yet available to benchmark the {\em achievable} precision in multimode multiparameter Gaussian quantum metrology.

This paper bridges such a gap. First, we develop a compact expression for the so-called quantum Fisher information matrix --- which determines the  precision available in quantum metrology through the quantum Cram\'er-Rao bound \cite{Helstrom1967,Braunstein1994,Matsumoto2002,PARIS2009} --- for multiparameter estimation with multimode Gaussian probes. This generalizes some partial instances from previous works \cite{Monras2010,Pinel12,Pinel13,Monras13,Gao2014,Jiang2014,Safranek2015,Safranek2016,Marian2016,Serafini2017} and provides a derivation independent from the fidelity formula obtained in \cite{Banchi15} by information geometry. Second, and most importantly, we also develop a compact formula to assess {\it compatibility} between pairs of parameters, that is, whether a common optimal measurement exists that allows one to estimate them jointly with minimum error \cite{Ragy2016}. This solves the problem of assessing the ultimate precision truly achievable in Gaussian multiparameter estimation, and provides a practical toolbox to validate effective metrological strategies for a variety of applications. These general results are presented in Sec.~\ref{sec:general}, after recalling the necessary preliminary notions in Sec.~\ref{sec:prelim}.

As an illustration, in Sec.~\ref{sec:example} we then consider the joint estimation of a phase shift and two noise parameters specifying a generic phase covariant Gaussian channel, using two-mode Gaussian probes in an interferometric setup. This extends previous studies where either phase only, or noise only, or phase and one noise parameter were estimated \cite{Caves1981,Monras2006,Monras07,AdessoAnno09,Gaiba09,Monras2011,Giovannetti2011,Pinel13,Gao2014,Gagatsos2016,Pirandola2017,Ragy2016}. We show that two-mode displaced squeezed probes with optimally tuned displacement and squeezing enable the simultaneous estimation of all three parameters, with an advantage over individual estimation that rapidly grows with increasing mean energy of the probes. We draw our concluding remarks in Sec.~\ref{sec:conclusions}

\section{Preliminaries}\label{sec:prelim}
\subsection{Gaussian states and Gaussian channels}
An $m$-mode bosonic CV system \cite{Braunstein2005,Weedbrook2012,Adesso2014,Serafini2017} is usually described in terms of a vector of quadrature operators $\hat{\textbf{R}}=\{\hat{q}_1,\hat{p}_1,\dots,\hat{q}_m,\hat{p}_m\}^\top$, which satisfy the canonical commutation relation \begin{equation}[\hat{R}_j,\hat{R}_k]=i\Omega_{jk},\end{equation} with $\Omega=i\sigma_y^{\oplus m}$. Here and in the following, $\mathbb{1},\sigma_x,\sigma_y,\sigma_z$ stand for the $2\times2$ identity and the Pauli matrices, respectively, and we adopt natural units ($\hbar =1$).
It is  convenient to describe the density matrix $\hat{\rho}$ of a CV system by its so-called characteristic function \cite{Ferraro2005,Serafini2017} \begin{equation}\chi_{\hat{\rho}}(\textbf{r})=\mbox{tr}[\hat{\rho}\hat{D}_{-\textbf{r}}]\,,\end{equation} where \begin{equation}\hat{D}_{-\textbf{r}}=e^{-i\textbf{r}^\top\Omega\hat{\textbf{R}}}\end{equation} is the displacement operator, and $\textbf{r}=\{q_1,p_1,\dots,q_m,p_m\}^\top$ is a vector of $2m$ real phase space coordinates.

An $m$-mode Gaussian state is a CV state with Gaussian characteristic function, \begin{equation}\label{eq:gaussianchi} \chi_{\hat{\rho}}(\textbf{r})=\exp\left[-\frac{1}{2}\textbf{r}^\top\Omega V\Omega\textbf{r}-i(\Omega\mathbf{d})^\top\textbf{r}\right]\,, \end{equation} and hence it is fully characterized by the first and second statistical moments of its quadrature operators, i.e., the displacement vector $\mathbf{d}=\langle\hat{\textbf{R}}\rangle$ and the covariance matrix $V$ with elements  \begin{equation} V_{jk}=\langle\{\hat{R}_j-d_j,\hat{R}_k-d_k\}_+\rangle\,, \end{equation} where $\{\cdot,\cdot\}_+$ is the anticommutator, and the uncertainty principle imposes
\begin{equation}V \geq i  \Omega \end{equation} for any physical state \cite{Simon1994}.
The mean energy per mode of an $m$-mode Gaussian state, i.e., the expectation value of the non-interacting quadratic Hamiltonian $\hat{H}=\omega\sum_{k=1}^m\left(\hat{a}_k^{\dagger} \hat{a}_k+1/2 \right)$ divided by the number of modes, with $\hat{a}_k = (\hat q_k + i \hat p_k)/\sqrt{2}$, can be easily computed from the covariance matrix $V$ and the displacement vector $\mathbf{d}$ of the state \cite{Weedbrook2012,Adesso2014}. In units of $\omega$, this is given by \begin{equation}\frac{\langle\hat{H}\rangle}{m}\equiv \bar{n}+\frac12=\frac{1}{2m}\left(\mbox{tr}~\frac{V}2+|\mathbf{d}|^2\right)\,,\end{equation} where $\bar{n}$ is the mean number of excitations per mode.

A Gaussian channel $\Lambda$ is a completely positive and trace-preserving map that transforms Gaussian states into Gaussian states \cite{Caves1994,Braunstein2005,Wolf2007,Weedbrook2012,Schaefer2013,Mari2014,Giovannetti2014,Serafini2017}. When a Gaussian channel preserves the number of modes of the input state, it can be represented (up to additional displacements) by two $2m\times2m$ real matrices, $X$ and  $Y$, with $Y=Y^\top$, which act on the displacement vector and the covariance matrix as
\begin{equation}
\begin{array}{l}
\mathbf{d}\rightarrow X\mathbf{d}, \\
V\rightarrow XVX^{\top}+Y, 
\end{array}
\end{equation}
and satisfy the complete positivity condition \begin{equation}Y+iX\Omega X^{\top}\geq i\Omega\,.\end{equation} The latter, for single-mode Gaussian channels, reads: $Y\ge0,\  \sqrt{\det Y}\geq|1-\det X|$. If the matrices representing the single-mode channel are proportional to the identity,
$X=\sqrt{x} \mathbb{1},~ Y=y \mathbb{1}$, with scalar parameters  $x,y \geq 0$, then the channel $\Lambda_{x,y}$ is said to be phase covariant
and the complete positivity condition reduces to $y \geq |1-x|$.


\subsection{Multiparameter quantum metrology}

In general, to implement an estimation protocol, one needs \cite{Giovannetti2006,PARIS2009,Giovannetti2011}: a probe state $\hat{\rho}_0$;  a physical mechanism described by a completely positive and trace-preserving map $\Lambda_{\{\mu\}}$ which encodes, on the probe state, the set of parameters $\{\mu\}$ one wishes to estimate; a measurement of the transformed state $\hat{\rho}_{\{\mu\}}=\Lambda_{\{\mu\}}[\hat{\rho}_0]$ and classical post-processing of the measurement results.
This procedure allows one to construct an estimator $\{\tilde{\mu}\}$ of the unknown parameters $\{\mu\}$, whose performance is quantified by the covariance matrix $\mbox{Cov}\left(\{\tilde\mu\}\right)$. Its diagonal elements, the variances, quantify the error in the estimation of the individual parameters, while the off-diagonal elements give an indication of the correlations between the parameters. The quantum Cram\'er-Rao bound yields a lower bound to the covariance matrix of an unbiased estimator in terms of the quantum Fisher information (QFI) matrix $\mathcal{F}$ \cite{Helstrom1967,Braunstein1994,Matsumoto2002,PARIS2009}:
\begin{equation}\label{eq:cramer-rao}
  \mbox{Cov}\left(\{\tilde{\mu}\}\right)\geq(M\mathcal{F})^{-1}~,
\end{equation}
where $M$ is the number of repetitions of the experiment.

In order to calculate the QFI matrix, one can introduce the symmetric logarithmic derivative (SLD) operators $\{\hat{\mathcal{L}}_\zeta\}_{\zeta\in\{\mu\}}$ which are implicitly defined by the equation 
\begin{equation}\hat{\mathcal{L}}_\zeta\hat\rho_{\{\mu\}}+\hat\rho_{\{\mu\}}\hat{\mathcal{L}}_\zeta =2\frac{\partial\hat\rho_{\{\mu\}}}{\partial\zeta}\,.\end{equation} 
These operators are hermitian, $\hat{\mathcal{L}}_\zeta=\hat{\mathcal{L}}_\zeta^\dagger$, by construction. The QFI matrix elements are then given by \begin{equation}
\mathcal{F}_{\eta\zeta}\equiv \mbox{$\frac{1}{2}$}\mbox{tr}\left(\hat{\rho}_{\{\mu\}}\{\hat{\mathcal{L}}_\eta,\hat{\mathcal{L}}_\zeta\}_+\right) =\Re\left[\mbox{tr}\left(\hat{\rho}_{\{\mu\}}\hat{\mathcal{L}}_\eta\hat{\mathcal{L}}_\zeta\right)\right]\,.\end{equation}


The Cram\'er-Rao bound, Eq.~(\ref{eq:cramer-rao}) can be saturated, in the limit $M \gg 1$ of an asymptotically large number of repetitions of the protocol, if an optimal measurement can be performed on the evolved state. For each parameter, an optimal measurement is described by a set of projectors which commute with its SLD. This implies that, if $[\hat{\mathcal{L}}_\eta,\hat{\mathcal{L}}_\zeta]=0$, then the existence of a common eigenbasis for the two SLDs is ensured, hence a jointly optimal measurement for extracting information on both parameters $\eta$ and $\zeta$ can be found. However, this condition is sufficient but not necessary. A weaker condition \cite{Matsumoto2002,Vaneph2013,Guta2013,Humphreys2013,Crowley2014,DattaReview,Ragy2016,Ciampini2017} states that the multiparameter Cram\'er-Rao bound can be asymptotically saturated iff  all pairs of SLDs commute ``on average'': (i) $\mathcal{J}_{\eta\zeta} = 0\ \ \forall\, \eta,\zeta\in\{\mu\}$, with
\begin{equation}\label{eq:jmunu}
\mathcal{J}_{\eta\zeta} \equiv \mbox{$\frac{1}{2i}$}\mbox{tr}\left(\hat{\rho}_{\{\mu\}}[\hat{\mathcal{L}}_\eta,\hat{\mathcal{L}}_\zeta]\right) =\Im\left[\mbox{tr}\left(\hat{\rho}_{\{\mu\}}\hat{\mathcal{L}}_\eta\hat{\mathcal{L}}_\zeta\right)\right]\,.
\end{equation}
Moreover, if one wishes to estimate each parameter as precisely as one would estimate them individually when assuming perfect knowledge of  the other parameters, then two more conditions need to be satisfied: (ii) there must exist a single probe state $\hat{\rho}_0$ that yields the optimal QFI for each of the parameters, and (iii) the parameters must be statistically independent, i.e., $\mathcal{F}_{\eta\zeta}=0~~\forall~\eta\neq\zeta$. The latter condition ensures that the uncertainty on one parameter does not affect the estimation precision of the others. When all conditions (i)--(iii) are met, then the parameters are said to be \emph{compatible} \cite{Ragy2016}.

Estimating $\kappa \equiv \left|\{\mu\}\right|$ parameters individually, where each parameter is estimated using the state $\hat{\rho}_0$, requires $\kappa$ times more resources (e.g.~energy, coherence or entanglement in the input state preparation \cite{Ragy2016,Braun2017}) than estimating them simultaneously using $\hat{\rho}_0$ in one shot\nocite{Humphreys2013}\nocite{Proctor2017}; however the latter strategy may not always offer superior performance if not all the parameters are compatible. A useful quantity to get a quantitative comparison of metrological performance between individual and simultaneous schemes is the ratio \cite{Yousefjani2017}
\begin{equation}\label{eq:ratio}
  \mathcal{R}=\frac{\Delta^{\mathrm{ind}}}{\Delta^{\mathrm{sim}}}~,
\end{equation}
where  (considering a single repetition, $M=1$):
\begin{equation}\label{eq:deltas}
\mbox{$\Delta^{\mathrm{ind}}=\sum_{\eta\in \{\mu\}}{{\mathcal{F}^{-1}_{\eta\eta}}} \mbox{\quad and \quad}
\Delta^{\mathrm{sim}}=\kappa^{-1}\:\mbox{tr}\,(\mathcal{F}^{-1})$}
\end{equation}
are the minimal total variances in the individual and simultaneous cases, respectively. Here the factor of $\kappa^{-1}$ is needed to account for the fact that the simultaneous scheme requires $\kappa$ less resources than individually estimating each parameter resetting the probe each time. When $\mathcal {R}>1$, simultaneous estimation is advantageous over individual, with a maximum advantage $\mathcal {R} = \kappa$ reachable only when condition (iii) holds.

\section{Gaussian quantum metrology} \label{sec:general}

Given an $m$-mode Gaussian state $\hat{\rho}_{\{\mu\}}$, depending on the set of parameters $\{\mu\}$, and denoting in what follows $\partial_\zeta\, \cdot \equiv \partial \cdot /\partial \zeta$, the SLD $\hat{\mathcal{L}}_\zeta$ for one of the parameters $\zeta \in \{\mu\}$ can be written as \cite{Serafini2017}
\begin{eqnarray}
  \hat{\mathcal{L}}_\zeta &\equiv& L^{(0)}_\zeta + \textbf{L}_\zeta^{(1)\top} \hat{\textbf{R}} + \hat{\textbf{R}}^{\top} \mbox{L}^{(2)}_{\zeta}\hat{\textbf{R}}\,,\label{eq:L} \\
\!\!\!\! \mbox{with:} \quad L_\zeta^{(0)}&=&-\mbox{$\frac{1}{2}$}\mbox{tr}\left(V_{\{\mu\}} \mbox{L}^{(2)}_\zeta\right)-\textbf{L}_\zeta^{(1)\top}\mathbf{d}_{\{\mu\}}- \mathbf{d}_{\{\mu\}}^{\top}\mbox{L}^{(2)}_\zeta\mathbf{d}_{\{\mu\}}\,,\label{eq:L0}\\
  \textbf{L}^{(1)}_\zeta&=&2V_{\{\mu\}}^{-1}\partial_\zeta \mathbf{d}_{\{\mu\}}-2\mbox{L}^{(2)}_\zeta\mathbf{d}_{\{\mu\}}\,,\label{eq:L1}\\
  \mbox{L}^{(2)}_\zeta&=& \sum_{j,k=1}^m \sum_{l=0}^{3} \frac{(a_{\zeta})^{jk}_l}{\nu_j\nu_k-(-1)^l}{S^{\top}}^{-1}\mbox{M}_l^{jk}S^{-1}\,.\label{eq:L2}
\end{eqnarray}
In the formulas above,  $\{\nu_i\}$ are the symplectic eigenvalues of the covariance matrix $V_{\{\mu\}}$, $S^{-1}$ is the symplectic transformation that brings $V_{\{\mu\}}$ into its diagonal form, $$S^{-1}V_{\{\mu\}}{S^{\top}}^{-1}=\boldsymbol{\nu}_{\{\mu\}}=\bigoplus_{i=1}^m \nu_i\mathds{1},$$ and
 $${(a_{\zeta})^{jk}_l}=\mbox{tr}\left(S^{-1}\partial_\zeta V_{\{\mu\}} {S^{\top}}^{-1}\mbox{M}_l^{jk}\right),$$ where the set of matrices $\mbox{M}_l^{jk}$ have all zero entries except for the $2 \times 2$ block in position $jk$ which is given by
$$ \big\{ M_l^{jk}\big\}_{l\in\{0,\dots,3\}} = \frac{1}{\sqrt{2}}\big\{i \sigma_y,~\sigma_z,~\mathds{1},~\sigma_x\big\}\,.
$$ 


As the main result of this paper, we obtain the following:

\begin{theorem}\label{thg}
  Given a CV bosonic Gaussian state of an arbitrary number of modes $m$, described by its first and second statistical moments $\mathbf{d}_{\{\mu\}}$ and $V_{\{\mu\}}$, respectively, and depending on the set of parameters $\{\mu\}$, we have for any $\eta,\zeta \in \{\mu\}$ that:
  \begin{eqnarray}
\!\!\!\!\!\!\!\!  \mathcal{F}_{\eta\zeta}&=&\mbox{$\frac{1}{2}$}\mathrm{tr}\left[(\partial_\zeta V_{\{\mu\}}) \mbox{\emph{L}}_\eta^{(2)}\right]+2(\partial_\eta\mathbf{d}_{\{\mu\}}^{\top})V_{\{\mu\}}^{-1}(\partial_\zeta\mathbf{d}_{\{\mu\}})\,, \label{eq:gqfim}\\
\!\!\!\!\!\!\!\!  \mathcal{J}_{\eta\zeta}&=&2\mathrm{tr}\left(\Omega \mbox{\emph{L}}_\zeta^{(2)}V_{\{\mu\}} \mbox{\emph{L}}_\eta^{(2)}\right) +2(\partial_\eta\mathbf{d}_{\{\mu\}}^{\top})V_{\{\mu\}}^{-1}\Omega V_{\{\mu\}}^{-1}(\partial_\zeta\mathbf{d}_{\{\mu\}})\,,\label{eq:gcompatibility}
  \end{eqnarray}
  with $\mathrm{L}_\zeta^{(2)}$ defined by Eq.~(\ref{eq:L2}).
\end{theorem}

Equation~(\ref{eq:gqfim}) provides a compact expression for the QFI matrix in Gaussian quantum metrology, directly generalizing the formula for the single-parameter case which can be found e.g.~in \cite{Serafini2017,Braun2017}.  Equation~(\ref{eq:gcompatibility}), on the other hand, provides a general formula for the quantity defined in Eq.~(\ref{eq:jmunu}), which determines the measurement compatibility condition (i) between pairs of parameters \cite{Ragy2016}. The proof of Theorem~\ref{thg} is given in Appendix~\ref{app:A}. Note that, while a formula equivalent to Eq.~(\ref{eq:gqfim}) may be alternatively derived from the expression for the quantum fidelity between two Gaussian states as recently reported in  \cite{Banchi15},  the formula in Eq.~(\ref{eq:gcompatibility}) is entirely original in the context of Gaussian quantum metrology and, to the best of our knowledge, no similar expression can be found in previous literature; in particular \cite{Banchiprivate}, Eq.~(\ref{eq:gcompatibility}) cannot be derived using the information geometry methods of \cite{Banchi15}.

Let us remark that both formulas appearing in Theorem~\ref{thg} can be evaluated efficiently for an arbitrary Gaussian state $\hat{\rho}_{\{\mu\}}$, although one needs to determine explicitly the symplectic transformation $S^{-1}$ that diagonalizes the covariance matrix $V_{\{\mu\}}$. The latter transformation can be constructed analytically for one and two modes, see e.g.~\cite{Serafini2005,Serafini2009,Mista2017}, and in general can be obtained numerically for a higher number of modes.

\section{Application to noisy optical interferometry}\label{sec:example}
\begin{figure}[t]
\centering
\includegraphics[width=7cm]{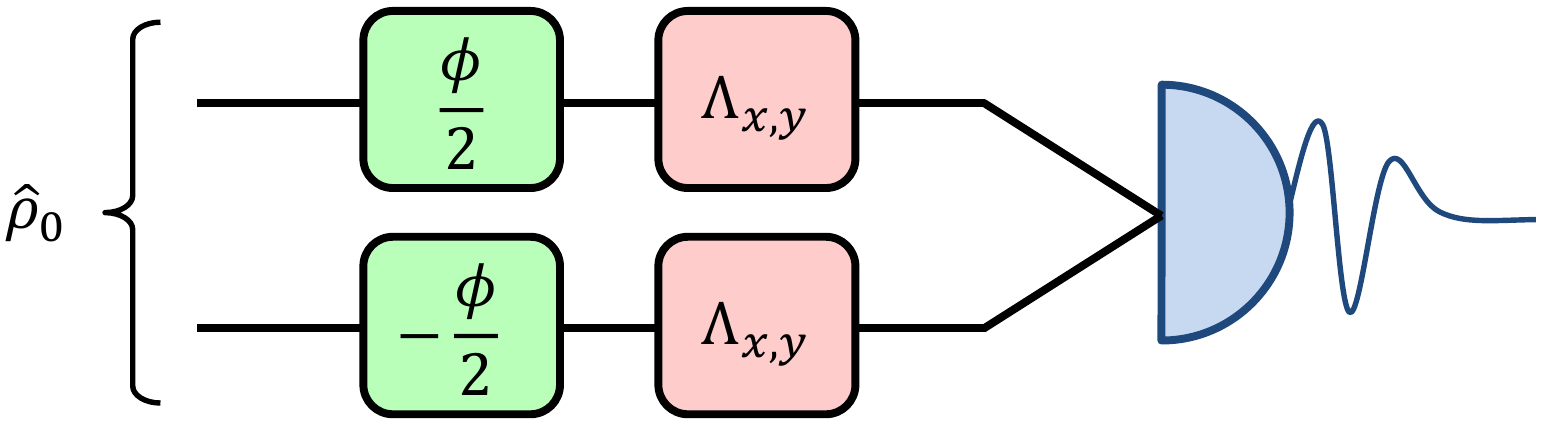}
\caption{(Color online) An instance of multiparameter Gaussian quantum metrology. The initial state $\hat{\rho}_0$ is a two-mode displaced squeezed state  which passes through an interferometric set-up before a joint measurement is made. One mode undergoes a phase transformation of $\phi/2$ and the other of $-\phi/2$, while both modes are affected by a phase covariant Gaussian channel $\Lambda_{x,y}$ with noise parameters $x$ and $y$. We determine optimal strategies for the estimation of the three parameters $\{\phi,x,y\}$.}
\label{fig:Scheme}
\end{figure}

To illustrate the usefulness of our results, we apply the general formalism of Theorem~\ref{thg} to the technologically relevant task of quantum phase estimation under noise in optical interferometry \cite{Caves1981,Huelga1997,Giovannetti2011,Escher2011,Rafau2012,DattaReview}. We focus specifically on the scheme of Fig.~\ref{fig:Scheme}, where an initial two-mode displaced squeezed state (TMDSS) $\hat{\rho}_0$ undergoes a phase transformation and transmission noise in an interferometric set-up, before the two modes are jointly measured. Here, we define the TMDSS as having:
\begin{eqnarray}\label{eq:tmss}
	\mathbf{d}_0 &=& \sqrt{2} \left\lbrace \Re\left[\alpha\right],  \Im\left[\alpha\right],  \Re\left[\beta\right],  \Im\left[\beta\right] \right\rbrace ^\top\,, \\
	V_0 &=&
	\left(
	\begin{array}{cccc}
	\cosh (2 r) & 0 & \sinh (2 r) & 0 \\
	0 & \cosh (2 r) & 0 & -\sinh (2 r) \\
	\sinh (2 r) & 0 & \cosh (2 r) & 0 \\
	0 & -\sinh (2 r) & 0 & \cosh (2 r) \\
	\end{array}
	\right)\,,\nonumber
\end{eqnarray}
where  $\alpha,\beta \in \mathds{C}$ are the displacements of each mode, and $r \in \mathds{R}$ is the squeezing parameter.
The phase difference $\phi$ is imprinted by each mode undergoing a unitary shift of $\pm\phi/2$, while the noise takes the form of a generic phase covariant Gaussian channel, $\Lambda_{x,y}$, on each mode. This includes the combined effect of loss ($0 \leq x  \leq 1$), amplification ($x \geq 1$), and added thermal noise ($y \geq |1-x|$), modelling realistic transmission of the probes in free space or over telecommunication fibres \cite{Caves1994,Braunstein2005,Wolf2007,Weedbrook2012,Schaefer2013,Mari2014,Giovannetti2014,Serafini2017}.  Our goal is to determine the best strategy to estimate all the three parameters $\phi$, $x$ and $y$ \footnote{Special instances of this problem have been considered in the literature, where the estimation of only one or two such parameters was addressed \cite{Caves1981,Monras2006,Monras07,AdessoAnno09,Gaiba09,Monras2011,Giovannetti2011,Pinel13,Gao2014,Gagatsos2016,Pirandola2017,Ragy2016}.},  hence tracking both signal ($\phi$) and noise ($x,y$) as precisely and efficiently as possible, using affordable TMDSS probes.

We first consider under which circumstances the compatibility condition (i) is obeyed, that is, when there exists a single optimal measurement for extracting all of the parameters such that the quantum Cram\'er-Rao bound (\ref{eq:cramer-rao}) may be asymptotically saturated. Using Eq.~(\ref{eq:gcompatibility}), one can easily show that this condition becomes dependent only on the displacement of the state, reading: $\left| \alpha \right|^2 = \left| \beta \right|^2 $. Further examining when a TMDSS  (\ref{eq:tmss}) leads to minimal total variances, $\Delta^{\mathrm{ind}}$ and  $\Delta^{\mathrm{sim}}$ [see Eq.~(\ref{eq:deltas})], we find that optimal states minimizing both of these quantities (assuming without loss of generality $r \geq 0$) must have $ \Re\left[\alpha\right] = \Re\left[\beta\right] = 0 $ and $ \Im\left[\alpha\right] = \Im\left[\beta\right]$. Consequently, any optimal input TMDSS already automatically obeys the measurement compatibility condition (i).

\begin{figure}[t]
\centering
\includegraphics[width=8.5cm]{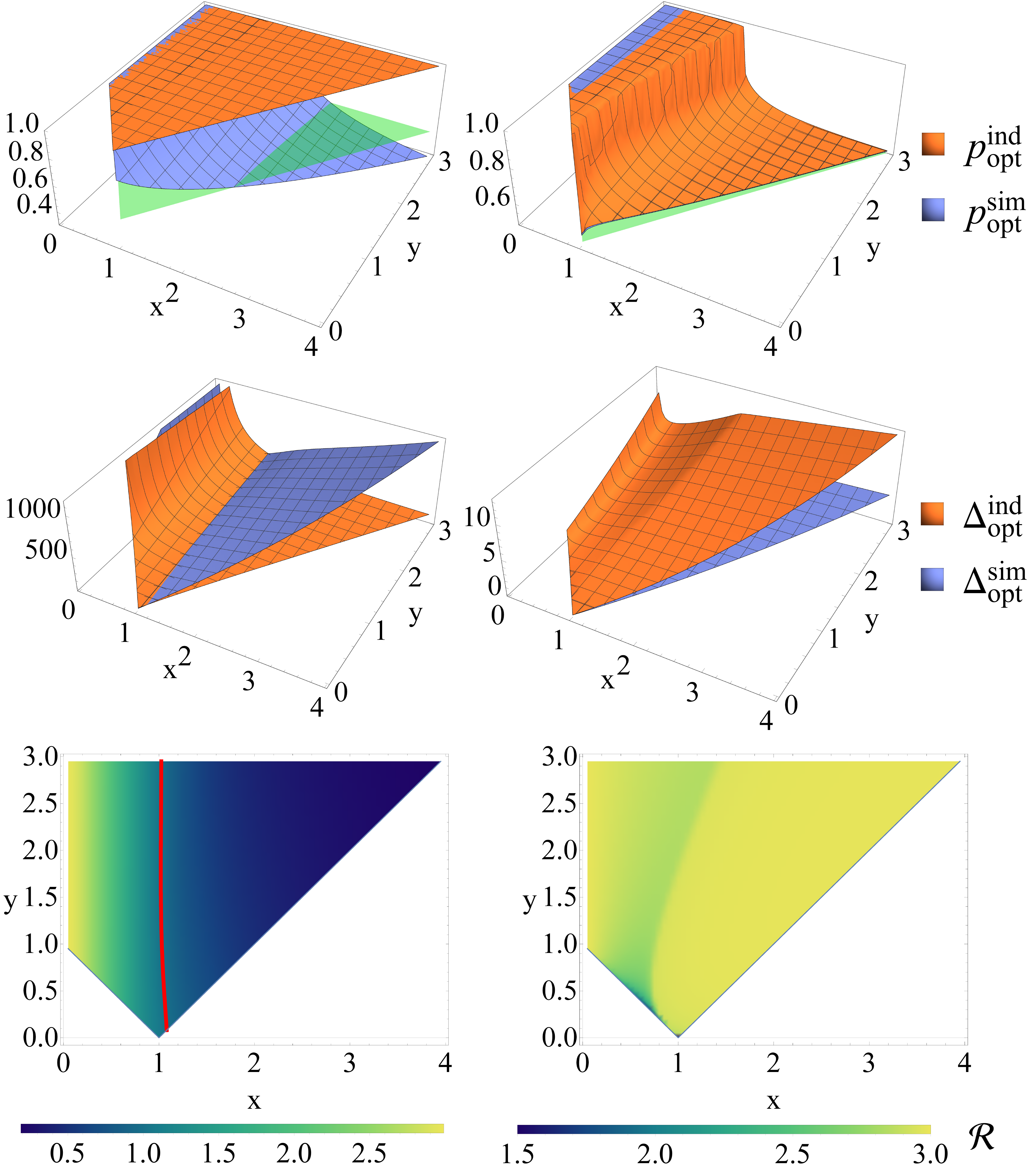}
\caption{(Color online)
	Results obtained from the metrological scheme outlined in Fig.~\ref{fig:Scheme}.
	Left column: Analysis at  low energy, $\bar n = 0.005$. Right column: Analysis at  higher energy,  $\bar n = 5$.
	Top: Optimal proportion of input energy to put in the displacement, $p_{\mathrm{opt}}$, comparing individual and simultaneous estimation schemes. The unmeshed (green online) plane marks $p_{\mathrm{opt}} = 1/2$, above which more energy should be used for displacement than squeezing.
	Middle: Minimal achievable error $\Delta_\mathrm{opt}$ for the two strategies.
	Bottom: Performance ratio ${\cal R} =  \Delta_\mathrm{opt}^\mathrm{ind}/\Delta_\mathrm{opt}^\mathrm{sim}$. The solid (red online) line marks ${\cal R}=1$;  when ${\cal R}>1$, the simultaneous estimation scheme outperforms the individual one.  All the presented results are independent of the value of the unknown phase $\phi$. All the quantities plotted are dimensionless.}
\label{fig:figures}
\end{figure}

We can then compare individual versus simultaneous estimation schemes at fixed input mean energy, customarily regarding the latter as the main resource for the metrological protocol \cite{PARIS2009,Giovannetti2011,Dowling2015,Toth2014}. The mean energy per mode of a TMDSS (\ref{eq:tmss}) with $\left| \alpha \right|^2 = \left| \beta \right|^2 $ 
can be written as $\langle \hat{H} \rangle/2 = \bar n + 1/2$, with  
\begin{equation}\bar n = \sinh^2(r) + |\alpha|^2\,.\end{equation} 

We find that the proportion $p \equiv |\alpha|^2/\bar n$ of this energy that is optimal to invest in displacement rather than squeezing varies with $x$ and $y$ (but not $\phi$) as well as with the total $\bar n$. Said optimal proportion $p_{\mathrm{opt}}$, in regimes of lower and higher input mean energy, is plotted in Fig.~\ref{fig:figures} (top) against the values of parameters $x$ and $y$. We observe that, at low energy, the individual and simultaneous schemes differ significantly. The optimal state for minimizing $\Delta_{\mathrm{ind}}$ has all of its energy dedicated to displacement, whereas for simultaneous estimation all energy should be put into displacement for $x \rightarrow 0$ and the optimal proportion $p_{\mathrm{opt}}$ decreases from 1 as $x$ increases, eventually dropping below $1/2$, in which case it is beneficial to put more energy in squeezing than displacement. As the available input energy $\bar n$ increases, this behavior quickly changes: the values of $p_\mathrm{opt}$ for the individual and simultaneous schemes become very similar. In both cases, it is always beneficial to put more energy into displacement than squeezing, with $p_\mathrm{opt}$ approaching, but never crossing, the plane $p_\mathrm{opt} = 1/2$. In both strategies, there is a region of parameters at low $x$ where the optimal input state has all its energy in displacement.

The middle and bottom rows of Fig.~\ref{fig:figures} both compare the minimal achievable total variances $\Delta_\mathrm{opt}^\mathrm{ind}$ and $\Delta_\mathrm{opt}^\mathrm{sim}$ for individual and simultaneous estimation, as defined in Eq.~(\ref{eq:deltas}) with $\kappa=3$ in our scheme. Specifically, the middle row shows how each variance changes with the noise parameters, while the bottom row illustrates the performance ratio $\mathcal{R}$ defined in Eq.~(\ref{eq:ratio}). At low  input energy, there are distinct regions of the parameter space  where either the individual estimation scheme ($\mathcal{R}<1$) or the simultaneous scheme ($\mathcal{R}>1$) is preferable. The solid (red) line in Fig.~\ref{fig:figures} (bottom, left) shows the boundary between these two regions. As the energy $\bar n$  approaches zero, this line approaches the vertical line $x=1$, i.e., the boundary between loss (where simultaneous estimation is superior) and amplification channels (where individual prevails instead). As the available energy $\bar n$ increases, this boundary moves quickly to the right, such that for any reasonable values of the parameters one gets simultaneous estimation as the optimal scheme. In fact, as Fig.~\ref{fig:figures} (bottom, right) shows, $\mathcal{R}$ then approaches its maximum value $3$ in a wide region of the parameter space. This shows that all three parameters become very nearly statistically independent, eventually fulfilling the compatibility condition (iii), with increasing  input energy $\bar n$ \footnote{Notice that ${\cal F}_{\phi x}={\cal F}_{\phi y}=0$, but  in general ${\cal F}_{xy}\neq 0$  \cite{Supp}.}. Quantitatively, by a series expansion we find that in the limit $\bar n \gg 0$ the ratio converges to
\begin{equation}{\cal R} \approx 3\left[1- \frac{\left(x^2+y^2+1\right)^2}{4 x^3 y} \bar{n}^{-3} + O(\bar{n}^{-4})\right]\,.
\end{equation}

We further observe that both individual and simultaneous total variances display at best a standard quantum limit scaling with the input energy, $\Delta_\mathrm{opt}^\mathrm{ind,sim} \lesssim O(\bar n^{-1})$, with no sub-shot-noise enhancement possible due to the presence of noise, in agreement with the general predictions of Refs.~\cite{Escher2011,Rafau2012,Pirandola2017}.
For completeness, the explicit QFI matrix $\cal F$ for the three parameters $\{\phi,x,y\}$, as computed from Eq.~(\ref{eq:gqfim}) using optimized TMDSS probes, is reported in Appendix~\ref{app:B}. We finally note that, in the individual estimation scheme, we have determined an optimal input state that minimizes the total variance $\Delta^{\mathrm{ind}}$ as defined in Eq.~(\ref{eq:deltas}), but one could in principle consider a different input state optimized independently for the estimation of each parameter. This analysis, reported in Appendix~\ref{app:B2}, leads to slighly better performances for individual estimation, but does not change any of the conclusions discussed above, including the fulfilment of the compatibility conditions and the qualitative regime where simultaneous estimation is advantageous.

\section{Conclusions}\label{sec:conclusions}
In this paper we derived general formulas to assess the ultimate precision available in Gaussian quantum metrology, that is, in the estimation of multiple parameters encoded in multimode Gaussian quantum states \cite{Serafini2017}.  We derived a compact expression, in terms of first and second moments of the states, for the quantum Fisher information matrix,  which bounds the achievable estimation error via the quantum Cram\'er-Rao bound. 
We then obtained 
a compact analytical expression to assess iff such a bound can be asymptotically saturated, i.e., iff a common measurement exists that is able to extract information optimally on all the parameters, a condition known as measurement compatibility \cite{Ragy2016}. This yields a general tool to endorse feasible estimation strategies in multiparameter Gaussian quantum metrology.

We applied our general formalism to study the practical estimation of three relevant physical parameters in noisy optical interferometry: an unknown phase shift and two unknown noise terms which specify a generic phase covariant Gaussian channel, capturing realistic instances of imperfect transmission. We showed that, using two-mode displaced squeezed input probes with optimally tuned squeezing and displacement, the measurement compatibility condition is satisfied, and one can estimate all three parameters simultaneously with an advantage over individual schemes growing rapidly as the available input energy is increased, eventually reaching a regime where the parameters are {\it de facto} statistically independent.

Our techniques can be promptly applied to a broad range of problems in fundamental science and technology  \cite{PARIS2009,Giovannetti2011,Dowling2015}, including (gravitational) interferometry, biosensing, imaging, positioning, thermometry, and more generally wherever the precise estimation of parameters encoded in quadratic Hamiltonians or noisy evolutions preserving Gaussianity is demanded. While this work focused mainly on compatibility conditions (i) and (iii), i.e.~measurement compatibility and statistical independence \cite{Ragy2016}, our framework can be combined with efficient numerical algorithms to find optimal input probe states \cite{Frowis2014,Knott2016,Safranek2016}, in order to fulfil condition (ii) and minimize the overall error on estimating multiple parameters. Tailoring existing algorithms --- or devising new ones --- to search within multimode Gaussian states, constrained to the compatibility constraints derived here, may be a valuable next step.

\begin{acknowledgments}
We acknowledge discussions with L. Banchi, D. Braun, M.G. Genoni, J. Liu, L. Mi\v{s}ta Jr., M.G.A. Paris, O. Pinel, S. Pirandola, D. \v{S}afr\'anek, A. Serafini, T. Tufarelli, and Q. Zhuang. This work was supported by the European Research Council  under the StG GQCOP (Grant No.~637352), the Royal Society under the International Exchanges
Programme (Grant No.~IE150570), and the Foundational Questions Institute under the Physics of the Observer Programme (Grant No.~FQXi-RFP-1601). P.K. acknowledges support from the Royal Commission for the Exhibition of 1851.
R.N. and P.L.-S. contributed equally to this work.

\emph{Note added} --- After submission of this work, an independent derivation of Eq.~(\ref{eq:gqfim}) was reported by D. \v{S}afr\'anek in \cite{Safranek2018}.

\end{acknowledgments}

\begin{widetext}
\appendix

\section{Proof of Theorem~1}\label{app:A}

To prove Theorem~\ref{thg} we will calculate explicitly, term by term, the following expression (adopting the notation
$\textbf{L}^{(1)}_\eta = ({L_{\eta\,l}^{(1)}})$
and
$\mbox{L}^{(2)}_\eta=({L_{\eta\,jk}^{(2)}})$, for $j,k,l\in\{1,\ldots,m\}$, and assuming here and in the following a sum over repeated indices):
\begin{equation*}
\mbox{tr}\left(\hat{\rho}_{\{\mu\}}\hat{\mathcal{L}}_\eta\hat{\mathcal{L}}_\zeta\right)= \mbox{tr}\left(\hat{\rho}_{\{\mu\}}\left(L^{(0)}_\eta+L^{(1)}_{\eta\,l}\hat{R}_l+ L^{(2)}_{\eta\,jk}\hat{R}_j\hat{R}_k\right)\left(L^{(0)}_\zeta+L^{(1)}_{\zeta\,m}\hat{R}_m+ L^{(2)}_{\zeta\,pq}\hat{R}_p\hat{R}_q\right)\right)~,
\end{equation*}
i.e.~we will find the explicit expressions for $\mbox{tr}(\hat{\rho}_{\{\mu\}}\hat{R}_p\hat{R}_q)$, $\mbox{tr}(\hat{\rho}_{\{\mu\}}\hat{R}_l\hat{R}_p\hat{R}_q)$ and $\mbox{tr}(\hat{\rho}_{\{\mu\}}\hat{R}_j\hat{R}_k\hat{R}_p\hat{R}_q)$; recall that the linear term is just the displacement vector: $\mbox{tr}(\hat{\rho}_{\{\mu\}}\hat{R}_l)=d_l$.

We will make use of some properties of the symmetrically ordered characteristic function $\chi(\mathbf{r})$. The first property is that the expectation value of an operator is equal to the characteristic function associated to it evaluated in $\textbf{r}=0$. The second one is that given any bounded operator $\hat{O}$, the following holds as a consequence of the Baker-Campbell-Hausdorff decomposition of the displacement operator:
\begin{equation}\label{eq:property1}
\chi_{\hat{O}\hat{R}_j}(\mathbf{r})=\mbox{Tr}[\hat{{D}}_{-\mathbf{r}} \hat{O}\hat{R}_j]= \left(-i\frac{\partial~}{\partial\tilde{r}_j}- \frac{1}{2}\Omega_{jk}\tilde{r}_k\right)\chi_{\hat{O}}(\mathbf{r})~,
\end{equation}
where $\tilde{\mathbf{r}}  =\Omega \mathbf{r}$. The last property we use reads:
\begin{equation}\label{eq:property2}
\frac{\partial~}{\partial\tilde{r}_m}\chi_{\hat{\rho}_{\{\mu\}}}= \frac{\partial~}{\partial\tilde{r}_m} e^{-\frac{1}{4}\tilde{r}_kV_{\{\mu\},kj}\tilde{r}_j+id_{\{\mu\},l}\tilde{r}_l}= i d_{\{\mu\},m}-\frac{1}{2}V_{\{\mu\},mj}\tilde{r}_j~,
\end{equation}
and it follows directly from the definition (\ref{eq:gaussianchi}) of Gaussian characteristic function:
\begin{equation}
\chi_{\hat{\rho}_{\{\mu\}}}(\textbf{r})=\exp\left[-\frac{1}{2}\textbf{r}^\top\Omega V_{\{\mu\}}\Omega\textbf{r}-i(\Omega\mathbf{d}_{\{\mu\}})^\top\textbf{r}\right]~.
\end{equation}
In the rest of the proof, for the sake of a lighter notation, we will indicate with $\chi \equiv \chi_{\hat{\rho}_{\{\mu\}}}$ the symmetrically ordered characteristic function of the Gaussian state $\hat{\rho}_{\{\mu\}}$ (if not specified otherwise) and we will write $\partial_j$ for $\frac{\partial~}{\partial\tilde{r}_j}$. Moreover, we will drop the explicit dependence on the set of parameters $\{\mu\}$ from  $\boldsymbol{d}_{\{\mu\}}$ and $V_{\{\mu\}}$, that is, we will use the shortcuts $d_j \equiv (\boldsymbol{d}_{\{\mu\}})_j \equiv {d}_{\{\mu\},j}$ and $V_{jk} \equiv (V_{\{\mu\}})_{jk} \equiv V_{\{\mu\},jk}$.

\subsubsection*{Quadratic term: $\mbox{tr}(\hat{\rho}_{\{\mu\}}\hat{R}_p\hat{R}_q)$}

Making use of property~\eqref{eq:property1} we get
\begin{eqnarray}
\mbox{tr}(\hat{\rho}_{\{\mu\}}\hat{R}_p\hat{R}_q)&=& \left(-i\partial_q-\frac{1}{2}\Omega_{qq'}\tilde{r}_{q'}\right)\chi_{\hat{\rho}_{\{\mu\}}\hat{R}_q}\bigg|_{\tilde{\mathbf{r}}=\mathbf{0}}= \left(-i\partial_q-\frac{1}{2}\Omega_{qq'}\tilde{r}_{q'}\right) \left(-i\partial_p-\frac{1}{2}\Omega_{pp'}\tilde{r}_{p'}\right)\chi\bigg|_{\tilde{\mathbf{r}}=\mathbf{0}}\nonumber\\
&=&\left[(-i)^2\partial_q\partial_p\chi+\frac{i}{2}\Omega_{pp'}\partial_q(\tilde{r}_{p'}\chi) +\frac{i}{2}\Omega_{qq'}\tilde{r}_{q'}\partial_p\chi+\frac{1}{4}\Omega_{qq'}\Omega_{pp'}\tilde{r}_{q'}\tilde{r}_{p'}\chi\right]_{\tilde{\mathbf{r}}=\mathbf{0}}\nonumber\\
&=&\left[-\partial_q\partial_p\chi+\frac{i}{2}\Omega_{pp'}(\delta_{qp'}\chi+\tilde{r}_{p'}\partial_q\chi)\right]_{\tilde{\mathbf{r}}=\mathbf{0}}= -\partial_q\partial_p\chi|_{\tilde{\mathbf{r}}=\mathbf{0}}+\frac{i}{2}\Omega_{pq}~,\label{eq:rr1}
\end{eqnarray}
and exploiting property~\eqref{eq:property2} we find
\begin{eqnarray}
\partial_q\partial_p\chi&=&\left(id_p-\frac{1}{2} V _{pp'}\tilde{r}_{p'}\right)\partial_q\chi+ \partial_q\left[\left(id_p-\frac{1}{2} V _{pp'}\tilde{r}_{p'}\right)\right]\chi\nonumber\\
&=&\left(id_p-\frac{1}{2} V _{pp'}\tilde{r}_{p'}\right)\left(id_q-\frac{1}{2} V _{qq'}\tilde{r}_{q'}\right)\chi- \frac{1}{2} V _{pp'}\delta_{qp'}\chi\nonumber\\
&=&\left(id_p-\frac{1}{2} V _{pp'}\tilde{r}_{p'}\right)\left(id_q-\frac{1}{2} V _{qq'}\tilde{r}_{q'}\right)\chi- \frac{1}{2} V _{pq}\chi~.
\end{eqnarray}
Evaluating this last expression in $\tilde{\mathbf{r}}=\mathbf{0}$ we get
\begin{equation}
\partial_q\partial_p\chi|_{\tilde{\mathbf{r}}=\mathbf{0}}=-d_pd_q-\frac{1}{2} V _{pq}~,
\end{equation}
and plugging this into Eq.~\eqref{eq:rr1} we finally obtain
\begin{equation}
\mbox{tr}(\hat{\rho}_{\{\mu\}}\hat{R}_p\hat{R}_q)=d_pd_q+\frac{1}{2}( V _{pq}+i\Omega_{pq})~.
\end{equation}

\subsubsection*{Cubic term: $\mbox{tr}(\hat{\rho}_{\{\mu\}}\hat{R}_l\hat{R}_p\hat{R}_q)$}
Applying property~\eqref{eq:property1} we write
\begin{eqnarray}
\mbox{tr}(\hat{\rho}_{\{\mu\}}\hat{R}_l\hat{R}_p\hat{R}_q)&=&\left(-i\partial_q-\frac{1}{2}\Omega_{qq'}\tilde{r}_{q'}\right) \left(-i\partial_p-\frac{1}{2}\Omega_{pp'}\tilde{r}_{p'}\right)\left(-i\partial_l-\frac{1}{2}\Omega_{ll'}\tilde{r}_{l'}\right)\chi\bigg|_{\tilde{\mathbf{r}}=\mathbf{0}}\nonumber\\
&=&\left[(-i)^3\partial_q\partial_p\partial_l\chi- \frac{(-i)^2}{2}\Omega_{ll'}\partial_q\partial_p(\tilde{r}_{l'}\chi)- \frac{i}{4}\Omega_{pp'}\Omega_{ll'}\partial_q(\tilde{r}_{p'}\tilde{r}_{l'}\chi) \right.\nonumber\\
&~&-\frac{(-i)^2}{2}\Omega_{pp'}\partial_q(\tilde{r}_{p'}\partial_l\chi)- \frac{(-i)^2}{2}\Omega_{qq'}\tilde{r}_{q'}\partial_p\partial_q\chi- \frac{i}{4}\Omega_{qq'}\Omega_{ll'}\tilde{r}_{q'}\partial_p(\tilde{r}_{l'}\chi)\nonumber\\
&~&\left.-\frac{i}{4}\Omega_{qq'}\Omega_{pp'}\tilde{r}_{q'}\tilde{r}_{p'}\partial_l\chi- \frac{1}{8}\Omega_{qq'}\Omega_{pp'}\Omega_{ll'}\tilde{r}_{q'}\tilde{r}_{p'}\tilde{r}_{l'}\chi\right]_{\tilde{\mathbf{r}}=\mathbf{0}}\nonumber\\
&=&\left[i\partial_q\partial_p\partial_l\chi+ \frac{1}{2}\Omega_{ll'}\partial_q\partial_p(\tilde{r}_{l'}\chi)+ \frac{1}{2}\Omega_{pp'}\partial_q(\tilde{r}_{p'}\partial_l\chi)\right]_{\tilde{\mathbf{r}}=\mathbf{0}}~.\label{eq:rrr1}
\end{eqnarray}
Making use of property~\eqref{eq:property2}, the three terms above are readily found. When evaluated in $\tilde{\mathbf{r}}=\mathbf{0}$, they read
\begin{eqnarray*}
\partial_q\partial_p(\tilde{r}_{l'}\chi)|_{\tilde{\mathbf{r}}=\mathbf{0}}&=&id_q\delta_{pl'}+id_p\delta_{ql'}~,\\
\partial_q(\tilde{r}_{p'}\partial_l\chi)|_{\tilde{\mathbf{r}}=\mathbf{0}}&=&id_l\delta_{qp'}~,\\
\partial_q\partial_p\partial_l\chi|_{\tilde{\mathbf{r}}=\mathbf{0}}&=&-id_pd_ld_q- \frac{i}{2}\left( V _{pl}d_q+ V _{pq}d_l+ V _{lq}d_p\right)~.
\end{eqnarray*}
Hence we get (notice that $ V _{jk}= V _{kj}$ as the covariance matrix is symmetric)
\begin{equation}
\mbox{tr}(\hat{\rho}_{\{\mu\}}\hat{R}_l\hat{R}_p\hat{R}_q)=d_pd_ld_q+ \frac{1}{2}\left[( V _{lp}+i\Omega_{lp})d_q+ ( V _{pq}+i\Omega_{pq})d_l+( V _{lq}+i\Omega_{lq})d_p\right]~.
\end{equation}

\subsubsection*{Quartic term: $\mbox{tr}(\hat{\rho}_{\{\mu\}}\hat{R}_j\hat{R}_k\hat{R}_p\hat{R}_q)$}

Considering that the linear term in $\tilde{r}_q$ gives no contribution when evaluated in $\tilde{\mathbf{r}}=\mathbf{0}$, we have
\begin{eqnarray}
\mbox{tr}\left(\hat{\rho}_{\{\mu\}}\hat{R}_j\hat{R}_k\hat{R}_p\hat{R}_q\right)&=&-i\partial_q \left(-i\partial_p-\frac{1}{2}\Omega_{pp'}\tilde{r}_{p'}\right) \left(-i\partial_k-\frac{1}{2}\Omega_{kk'}\tilde{r}_{k'}\right) \left(-i\partial_j-\frac{1}{2}\Omega_{jj'}\tilde{r}_{j'}\right) \chi\bigg|_{\tilde{\mathbf{r}}=\mathbf{0}}\nonumber\\
&=&-i\partial_q\left[(-i)^3\partial_p\partial_k\partial_j\chi- \frac{(-i)^2}{2}\Omega_{jj'}\partial_p\partial_k(\tilde{r}_{j'}\chi)- \frac{i}{4}\Omega_{kk'}\Omega_{jj'}\partial_p(\tilde{r}_{k'}\tilde{r}_{j'}\chi) \right.\nonumber\\
&~&\qquad\quad-\frac{(-i)^2}{2}\Omega_{kk'}\partial_p(\tilde{r}_{k'}\partial_j\chi)- \frac{(-i)^2}{2}\Omega_{pp'}\tilde{r}_{p'}\partial_k\partial_j\chi- \frac{i}{4}\Omega_{pp'}\Omega_{jj'}\tilde{r}_{p'}\partial_k(\tilde{r}_{j'}\chi)\nonumber\\
&~&\qquad\quad\left.-\frac{i}{4}\Omega_{pp'}\Omega_{kk'}\tilde{r}_{p'}\tilde{r}_{k'}\partial_j\chi- \frac{1}{8}\Omega_{pp'}\Omega_{kk'}\Omega_{jj'}\tilde{r}_{p'}\tilde{r}_{k'}\tilde{r}_{j'}\chi\right]_{\tilde{\mathbf{r}}=\mathbf{0}}~.\nonumber
\end{eqnarray}
Notice that the last two terms do not contribute when evaluated in $\tilde{\mathbf{r}}=\mathbf{0}$, since they are of the second and third order in $\tilde{r}$ and, when differentiated with respect to $\tilde{r}_q$, they produce a linear and a quadratic term, respectively. We then get
\begin{eqnarray}
\mbox{tr}(\hat{\rho}_{\{\mu\}}\hat{R}_j\hat{R}_k\hat{R}_p\hat{R}_q)&=&\left[(-i)^4\partial_q\partial_p\partial_k\partial_j\chi -\frac{(-i)^3}{2}\Omega_{jj'}\partial_q\partial_p\partial_k(\tilde{r}_{j'}\chi)\right.\nonumber\\
&~&+ \frac{(-i)^2}{4}\Omega_{kk'}\Omega_{jj'}\partial_q\partial_p(\tilde{r}_{k'}\tilde{r}_{j'}\chi)- \frac{(-i)^3}{2}\Omega_{kk'}\partial_q\partial_p(\tilde{r}_{k'}\partial_j\chi) \nonumber\\
&~&\left.- \frac{(-i)^3}{2}\Omega_{pp'}\partial_q(\tilde{r}_{p'}\partial_k\partial_j\chi)+ \frac{(-i)^2}{4}\Omega_{pp'}\Omega_{jj'}\partial_q\left(\tilde{r}_{p'}\partial_k(\tilde{r}_{j'}\chi)\right)\right]_{\tilde{\mathbf{r}}=\mathbf{0}}~.\label{eq:rrrr1}
\end{eqnarray}

The six terms above, when evaluated in $\tilde{\mathbf{r}}=\mathbf{0}$, give
\begin{eqnarray*}
\partial_q\partial_p\partial_k(\tilde{r}_{j'}\chi)|_{\tilde{\mathbf{r}}=\mathbf{0}}&=&\delta_{kj'}\partial_q\partial_p\chi|_{\mathbf{r}=\mathbf{0}}+ \delta_{pj'}\partial_q\partial_k\chi|_{\mathbf{r}=\mathbf{0}}+ \delta_{qj'}\partial_p\partial_k\chi|_{\mathbf{r}=\mathbf{0}}~,\\
\partial_q\partial_p(\tilde{r}_{k'}\tilde{r}_{j'}\chi)|_{\tilde{\mathbf{r}}=\mathbf{0}}
&=&\delta_{pk'}\delta_{qj'}+\delta_{pj'}\delta_{qk'}~,\\
\partial_q\partial_p(r_{k'}\partial_j\chi)|_{\tilde{\mathbf{r}}=\mathbf{0}}&=& \delta_{pk'}\partial_q\partial_j\chi|_{\mathbf{r}=\mathbf{0}}+\delta_{qk'}\partial_p\partial_j\chi|_{\mathbf{r}=\mathbf{0}}~,\\
\partial_q(\tilde{r}_{p'}\partial_k\partial_j\chi)|_{\tilde{\mathbf{r}}=\mathbf{0}}&=&\delta_{qp'}\partial_k\partial_j\chi|_{\mathbf{r}=\mathbf{0}}~,\\
\partial_q\left(\tilde{r}_{p'}\partial_k(\tilde{r}_{j'}\chi)\right)|_{\tilde{\mathbf{r}}=\mathbf{0}}
&=&\delta_{qp'}\delta_{kj'}~,\\
\partial_q\partial_p\partial_k\partial_j\chi|_{\tilde{\mathbf{r}}=\mathbf{0}}&=& d_kd_jd_qd_p+\frac{1}{2}\left( V _{qp}d_kd_j+ V _{kj}d_pd_q+ V _{kq}d_jd_p+ V _{jq}d_kd_p+  V _{kp}d_jd_q+ V _{jp}d_kd_q\right)\nonumber\\
&~&+\frac{1}{4}\left( V _{kp} V _{jq}+ V _{jp} V _{kq}+ V _{kj} V _{qp}\right)~.
\end{eqnarray*}
Plugging these expressions into Eq.~\eqref{eq:rrrr1}, we get
\begin{eqnarray}
\mbox{tr}(\hat{\rho}_{\{\mu\}}\hat{R}_j\hat{R}_k\hat{R}_p\hat{R}_q)&=&d_j d_k d_p d_q+ \frac{1}{2} d_p d_q  V  _{j k}+\frac{1}{2} d_k d_q  V _{j p}+\frac{1}{2} d_j d_q  V  _{k p}+\frac{1}{2} d_k d_p  V  _{j q} +\frac{1}{2} d_j d_p  V  _{k q}+\frac{1}{2} d_j d_k
V  _{p q}\nonumber\\
&~&+\frac{i}{2} \left\{\Omega _{j k} \left(d_p d_q+\frac{ V  _{p q}}{2}\right)+\Omega _{j p} \left(d_k d_q+\frac{ V  _{k q}}{2}\right)+\Omega _{k p} \left(d_j d_q+\frac{ V  _{j
		q}}{2}\right)\right.\nonumber\\
&~&\qquad\left.+\Omega _{j q} \left(d_k d_p+\frac{ V  _{k p}}{2}\right)+\Omega _{k q} \left(d_j d_p+\frac{ V  _{j p}}{2}\right)+\Omega _{p q} \left(d_j d_k+\frac{ V  _{j
		k}}{2}\right)\right\}\nonumber\\
&~&-\frac{1}{4} \left(\Omega _{j q} \Omega _{k p}+\Omega _{j p} \Omega _{k q}+\Omega _{j k} \Omega _{p q}\right)+\frac{1}{4}(  V  _{j q}  V  _{k p}+ V  _{j p}  V  _{k q}+  V  _{j k}  V  _{p q})~.
\end{eqnarray}

Before moving to the last part of the proof, we recall that the expectation value of the SLD operator is zero. This is easy to check:
\begin{eqnarray*}
\langle\hat{\mathcal{L}}_\zeta\rangle&=& \mbox{tr}\left(\hat{\rho}_{\{\mu\}}\hat{\mathcal{L}}_\zeta\right)= L_\zeta^{(0)}\mbox{tr}(\hat{\rho}_{\{\mu\}})+ L_{\zeta\,m}^{(1)}\mbox{tr}(\hat{\rho}_{\{\mu\}}\hat{R}_m)+ L_{\zeta\,jk}^{(2)}\mbox{tr}(\hat{\rho}_{\{\mu\}}\hat{R}_j\hat{R}_k)=\nonumber\\
&=&L_\zeta^{(0)}+L_{\zeta\,m}^{(1)}d_m+ L_{\zeta\,jk}^{(2)}\left(d_jd_k+\frac{1}{2}( V _{jk}+i\Omega_{jk})\right)~,
\end{eqnarray*}
which in vectorial form reads
\begin{eqnarray}\label{eq:avgL}
\langle\hat{\mathcal{L}}_\zeta\rangle&=& L_\zeta^{(0)}+\textbf{L}_\zeta^{(1)\mathsf{T}}\textbf{d}+ \textbf{d}^{\mathsf{T}}\mbox{L}_\zeta^{(2)}\textbf{d}+ \frac{1}{2}\mbox{tr}(\mbox{L}_\zeta^{(2)} V )+ \frac{i}{2}\mbox{tr}(\mbox{L}_\zeta^{(2)}\Omega)~.
\end{eqnarray}
When substituting the definition of $ L_\zeta^{(0)}$ from Eq.~\eqref{eq:L0} into Eq.~(\ref{eq:avgL}), we are left with a term proportional to $\mbox{tr}(\mbox{L}_\zeta^{(2)}\Omega)$, which vanishes because
$\mbox{L}_\zeta^{(2)}$ is symmetric while $\Omega$ is skew-symmetric.

\subsubsection*{Expressions for $\mbox{tr}\left(\hat{\rho}_{\{\mu\}}\hat{\mathcal{L}}_\eta\hat{\mathcal{L}}_\zeta\right)$, $\mathcal{F}_{\eta\zeta}$, $\mathcal{J}_{\eta\zeta}$}

We have that
\begin{eqnarray}
\mbox{tr}\left(\hat{\rho}_{\{\mu\}}\hat{\mathcal{L}}_\eta\hat{\mathcal{L}}_\zeta\right)&=& L_\eta^{(0)}L_{\zeta}^{(0)}+L_\eta^{(0)}L_{\zeta\,m}^{(1)}\mbox{tr}(\hat{\rho}_{\{\mu\}}\hat{R}_m)+ L_\eta^{(0)}L_{\zeta\,pq}^{(2)}\mbox{tr}(\hat{\rho}_{\{\mu\}}\hat{R}_p\hat{R}_q)\nonumber\\
&~&+L_{\eta\,l}^{(1)}L_\zeta^{(0)}\mbox{tr}(\hat{\rho}_{\{\mu\}}\hat{R}_l)+ L_{\eta\,l}^{(1)}L_{\zeta\,m}^{(1)}\mbox{tr}(\hat{\rho}_{\{\mu\}}\hat{R}_l\hat{R}_m)+ L_{\eta\,l}^{(1)}L_{\zeta\,pq}^{(2)}\mbox{tr}(\hat{\rho}_{\{\mu\}}\hat{R}_l\hat{R}_p\hat{R}_q)\nonumber\\
&~&+L_{\eta\,jk}^{(2)}L_\zeta^{(0)}\mbox{tr}(\hat{\rho}_{\{\mu\}}\hat{R}_j\hat{R}_k)+ L_{\eta\,jk}^{(2)}L_{\zeta\,m}^{(1)}\mbox{tr}(\hat{\rho}_{\{\mu\}}\hat{R}_j\hat{R}_k\hat{R}_m)+ L_{\eta\,jk}^{(2)}L_{\zeta\,pq}^{(2)}\mbox{tr}(\hat{\rho}_{\{\mu\}}\hat{R}_j\hat{R}_k\hat{R}_p\hat{R}_q)~,\nonumber
\end{eqnarray}
and exploiting the results of the previous sections we get
\begin{eqnarray}
\mbox{tr}\left(\hat{\rho}_{\{\mu\}}\hat{\mathcal{L}}_\eta\hat{\mathcal{L}}_\zeta\right)
&=&L_\eta^{(0)}L_{\zeta}^{(0)}+L_\eta^{(0)}L_{\zeta\,m}^{(1)}d_m+ L_\eta^{(0)}L_{\zeta\,pq}^{(2)}\left(d_pd_q+\frac{1}{2}( V _{pq}+i\Omega_{pq})\right)+ L_{\eta\,l}^{(1)}L_\zeta^{(0)}d_l+ L_{\eta\,l}^{(1)}L_{\zeta\,m}^{(1)}\left(d_ld_m+\frac{1}{2}( V _{lm}+i\Omega_{lm})\right)\nonumber\\
&~&+L_{\eta\,l}^{(1)}L_{\zeta\,pq}^{(2)}\left(d_pd_ld_q+ ( V _{lp}+i\Omega_{lp})\frac{d_q}{2}+ ( V _{pq}+i\Omega_{pq})\frac{d_l}{2}+ ( V _{lq}+i\Omega_{lq})\frac{d_p}{2}\right)
+L_{\eta\,jk}^{(2)}L_\zeta^{(0)}\left(d_jd_k+\frac{1}{2}( V _{jk}+i\Omega_{jk})\right)\nonumber\\
&~&+L_{\eta\,jk}^{(2)}L_{\zeta\,m}^{(1)}\left(d_jd_kd_m+ ( V _{jk}+i\Omega_{jk})\frac{d_m}{2}+ V _{km}+i\Omega_{km})\frac{d_j}{2}+( V _{j}+i\Omega_{jm})\frac{d_k}{2}\right)\nonumber\\
&~&+L_{\eta\,jk}^{(2)}L_{\zeta\,pq}^{(2)}\left\{d_j d_k d_p d_q+ \frac{1}{2} d_p d_q  V  _{j k}+\frac{1}{2} d_k d_q  V _{j p}+\frac{1}{2} d_j d_q  V  _{k p}\right.\nonumber\\
&~&\qquad\qquad\quad+\frac{1}{2} d_k d_p  V  _{j q}+\frac{1}{2} d_j d_p V  _{k q}+\frac{1}{2} d_j d_k
V  _{p q}+\frac{1}{4}  V  _{j q}  V  _{k p}+\frac{1}{4}  V  _{j p}  V  _{k q}+\frac{1}{4}  V  _{j k}  V  _{p q}\nonumber\\
&~&\qquad\qquad\quad+\frac{1}{2} i \left[\Omega _{j k} \left(d_p d_q+\frac{ V  _{p q}}{2}\right)+\Omega _{j p} \left(d_k d_q+\frac{ V  _{k q}}{2}\right)+\Omega _{k p} \left(d_j d_q+\frac{ V  _{j
		q}}{2}\right)+\Omega _{j q} \left(d_k d_p+\frac{ V  _{k p}}{2}\right)\right.\nonumber\\
&~&\qquad\qquad\qquad\quad\left.\left.+\Omega _{k q} \left(d_j d_p+\frac{ V  _{j p}}{2}\right)+\Omega _{p q} \left(d_j d_k+\frac{ V  _{j
		k}}{2}\right)\right]-\frac{1}{4} \left(\Omega _{j q} \Omega _{k p}+\Omega _{j p} \Omega _{k q}+\Omega _{j k} \Omega _{p q}\right)\right\} \nonumber \\
&=&\frac{1}{2} d_k  V  _{j m} L_{\eta\,jk}^{(2)} L_{\zeta\,m}^{(1)}+
\frac{1}{2} d_j L_{\eta\,jk}^{(2)}  V  _{k m} L_{\zeta\,m}^{(1)}
+\frac{1}{2} i d_k \Omega_{jm} L_{\eta\,jk}^{(2)} L_{\zeta\,m}^{(1)}+
\frac{1}{2} i d_j L_{\eta\,jk}^{(2)} \Omega_{km} L_{\zeta\,m}^{(1)}\nonumber\\
&~&
+\frac{1}{2} d_k d_q  V  _{j p}
L_{\eta\,jk}^{(2)} L_{\zeta\,pq}^{(2)}+\frac{1}{2} d_j d_q L_{\eta\,jk}^{(2)}  V  _{k p} L_{\zeta\,pq}^{(2)}
+\frac{1}{2} d_k d_p  V  _{j q} L_{\eta\,jk}^{(2)}L_{\zeta\,pq}^{(2)}+
\frac{1}{2} d_j d_p L_{\eta\,jk}^{(2)}  V  _{k q} L_{\zeta\,pq}^{(2)}\nonumber\\
&~&
+\frac{1}{2} i d_k d_q \Omega _{jp}
L_{\eta\,jk}^{(2)}L_{\zeta\,pq}^{(2)}+
\frac{1}{2} i d_k d_p \Omega _{jq} L_{\eta\,jk}^{(2)} L_{\zeta\,pq}^{(2)}
+\frac{1}{2} i d_j d_q L_{\eta\,jk}^{(2)} \Omega _{kp} L_{\zeta\,pq}^{(2)}+
\frac{1}{2} i d_j d_p L_{\eta\,jk}^{(2)} \Omega _{kq} L_{\zeta\,pq}^{(2)}\nonumber\\
&~&
+\frac{1}{2} d_q
L_{\eta\,l}^{(1)}  V  _{l p} L_{\zeta\,pq}^{(2)}+
\frac{1}{2} d_p L_{\eta\,l}^{(1)}  V  _{l q} L_{\zeta\,pq}^{(2)}
+\frac{1}{2} i d_q L_{\eta\,l}^{(1)} \Omega_{lp} L_{\zeta\,pq}^{(2)}+
\frac{1}{2} i d_p L_{\eta\,l}^{(1)} \Omega _{lq} L_{\zeta\,pq}^{(2)}\nonumber\\
&~&
+\frac{1}{4} i \Omega _{jp} L_{\eta\,jk}^{(2)}  V  _{k q} L_{\zeta\,pq}^{(2)}+
\frac{1}{4} i \Omega _{jq} L_{\eta\,jk}^{(2)}  V  _{k p} L_{\zeta\,pq}^{(2)}
+\frac{1}{4} i
V  _{j q} L_{\eta\,jk}^{(2)} \Omega _{kp} L_{\zeta\,pq}^{(2)}+
\frac{1}{4} i  V  _{j p} L_{\eta\,jk}^{(2)} \Omega _{kq} L_{\zeta\,pq}^{(2)}\nonumber\\
&~&
+\frac{1}{4}  V  _{j q} L_{\eta\,jk}^{(2)}  V  _{k p} L_{\zeta\,pq}^{(2)}+
\frac{1}{4}  V  _{j p} L_{\eta\,jk}^{(2)}  V  _{k q}
L_{\zeta\,pq}^{(2)}-\frac{1}{4} \Omega _{jq} L_{\eta\,jk}^{(2)} \Omega _{kp} L_{\zeta\,pq}^{(2)}-
\frac{1}{4} \Omega _{jp} L_{\eta\,jk}^{(2)} \Omega _{kq} L_{\zeta\,pq}^{(2)}\nonumber\\
&~&
+\frac{1}{2} L_{\eta\,l}^{(1)}  V  _{l m} L_{\zeta\,m}^{(1)}+
\frac{1}{2} i L_{\eta\,l}^{(1)} \Omega_{lm} L_{\zeta\,m}^{(1)}~,\label{eq:horror}
\end{eqnarray}
where to obtain the last equality we used the trick of subtracting $\langle\hat{\mathcal{L}}_\zeta\rangle\langle\hat{\mathcal{L}}_\eta\rangle = 0$.

In vectorial form, Eq.~(\ref{eq:horror}) becomes
\begin{eqnarray}
\mbox{tr}\left(\hat{\rho}_{\{\mu\}}\hat{\mathcal{L}}_\eta\hat{\mathcal{L}}_\zeta\right)
&=&\textbf{d}^{\mathsf{T}}\mbox{L}_\eta^{(2)} V \textbf{L}_\zeta^{(1)}+ i\textbf{d}^{\mathsf{T}}\mbox{L}_\eta^{(2)}\Omega\textbf{L}_\zeta^{(1)}+ 2\textbf{d}^{\mathsf{T}}\mbox{L}_\eta^{(2)} V  \mbox{L}_\zeta^{(2)}\textbf{d}+2i\textbf{d}^{\mathsf{T}}\mbox{L}_\eta^{(2)}\Omega L_\zeta\textbf{d}+ \textbf{d}^{\mathsf{T}}\mbox{L}_\zeta^{(2)} V \textbf{L}_\eta^{(1)}+ i\textbf{L}_\eta^{(1)\mathsf{T}}\Omega \mbox{L}_\zeta^{(2)}\textbf{d}\nonumber\\
&~&+2i\mbox{tr}\left(\Omega L_\zeta^{(2)} V  \mbox{L}_\eta^{(2)}\right)+ \frac{1}{2}\mbox{tr}\left( V  \mbox{L}_\zeta^{(2)} V  \mbox{L}_\eta^{(2)}\right)+ \frac{1}{2}\mbox{tr}\left(\Omega \mbox{L}_\zeta^{(2)}\Omega \mbox{L}_\eta^{(2)}\right)+\frac{1}{2}\textbf{L}_\eta^{(1)} V \textbf{L}_\zeta^{(1)}+ \frac{i}{2}\textbf{L}_\eta^{(1)\mathsf{T}}\Omega\textbf{L}_\zeta^{(1)}~.
\end{eqnarray}



Now, since for any two hermitian operators $\hat{A}$ and $\hat{B}$ it holds that
$2\mbox{tr}\left(\hat{\rho}_{\{\mu\}}\hat{A}\hat{B}\right)= \mbox{tr}\left(\hat{\rho}_{\{\mu\}}\{\hat{A},\hat{B}\}_+\right)+ \mbox{tr}\left(\hat{\rho}_{\{\mu\}}[\hat{A},\hat{B}]\right)$,
we find
\begin{eqnarray}
\Re\left\{\mbox{tr}\left(\hat{\rho}_{\{\mu\}}\hat{\mathcal{L}}_\eta\hat{\mathcal{L}}_\zeta\right)\right\}&=& \frac{1}{2}\mbox{tr}\left(\hat{\rho}_{\{\mu\}}\{\hat{\mathcal{L}}_\eta,\hat{\mathcal{L}}_\zeta\}_+\right)= \mathcal{F}_{\eta\zeta}=\mathcal{F}_{\zeta\eta}~,\label{eq:qfi} \\
\Im\left\{\mbox{tr}\left(\hat{\rho}_{\{\mu\}}\hat{\mathcal{L}}_\eta\hat{\mathcal{L}}_\zeta\right)\right\} &=& \frac{1}{2i}\mbox{tr}\left(\hat{\rho}_{\{\mu\}}[\hat{\mathcal{L}}_\eta,\hat{\mathcal{L}}_\zeta]\right)= \mathcal{J}_{\eta\zeta}=-\mathcal{J}_{\zeta\eta}~. \label{eq:j}
\end{eqnarray}

Using the cyclic property of the trace and the identity  $\partial_\zeta V = V  \mbox{L}_\zeta^{(2)} V +\Omega \mbox{L}_\zeta^{(2)}\Omega$ \cite{Serafini2017}, we have that
\begin{equation}
\frac{1}{2}\mbox{tr}\left( V  \mbox{L}_\zeta^{(2)} V  \mbox{L}_\eta^{(2)}\right)+ \frac{1}{2}\mbox{tr}\left(\Omega \mbox{L}_\zeta^{(2)}\Omega \mbox{L}_\eta^{(2)}\right)=\mbox{tr}\left(\partial_\zeta V  \mbox{L}_\eta^{(2)}\right)=\mbox{tr}\left(\partial_\eta V  \mbox{L}_\zeta^{(2)}\right)~,
\end{equation}
therefore, for Eq.~\eqref{eq:qfi}, we can write
\begin{eqnarray}
\Re\left\{\mbox{tr}\left(\hat{\rho}_{\{\mu\}}\hat{\mathcal{L}}_\eta\hat{\mathcal{L}}_\zeta\right)\right\}&=& \textbf{d}^{\mathsf{T}}\mbox{L}_\eta^{(2)} V \textbf{L}_\zeta^{(1)}+ 2\textbf{d}^{\mathsf{T}}\mbox{L}_\eta^{(2)} V  L_\zeta^{(2)}\textbf{d}+ \textbf{d}^{\mathsf{T}}\mbox{L}_\zeta^{(2)} V \textbf{L}_\eta^{(1)}+\frac{1}{2}\mbox{tr}(\partial_\zeta V  \mbox{L}_\eta^{(2)})+ \frac{1}{2}\textbf{L}_\eta^{(1)} V \textbf{L}_\zeta^{(1)}~.
\end{eqnarray}
Finally substituting in the expression for $\textbf{L}^{(1)}$ given by Eq.~\eqref{eq:L1}, and adopting in what follows the shorthand notation $\boldsymbol{d}_{\zeta} \equiv \partial_{\zeta}\textbf{d}$, we get
\begin{eqnarray}
\Re\left\{\mbox{tr}\left(\hat{\rho}_{\{\mu\}}\hat{\mathcal{L}}_\eta\hat{\mathcal{L}}_\zeta\right)\right\}&=& 2\textbf{d}^{\mathsf{T}}\mbox{L}_\eta^{(2)}\boldsymbol{d}_\zeta- 2\textbf{d}^{\mathsf{T}}\mbox{L}_\eta^{(2)} V  L_\zeta^{(2)}\textbf{d}+ 2\textbf{d}^{\mathsf{T}}\mbox{L}_\eta^{(2)} V  \mbox{L}_\zeta^{(2)}\textbf{d}+2\textbf{d}^{\mathsf{T}}\mbox{L}_\zeta^{(2)}\boldsymbol{d}_\eta- 2\textbf{d}^{\mathsf{T}}\mbox{L}_\zeta^{(2)} V  \mbox{L}_\eta^{(2)}\textbf{d}+ \frac{1}{2}\mbox{tr}(\partial_\zeta V  \mbox{L}_\eta^{(2)})\nonumber\\
&~&+2\boldsymbol{d}_\eta^{\mathsf{T}} V ^{-1}\boldsymbol{d}_\zeta- 2\boldsymbol{d}_\eta^{\mathsf{T}}\mbox{L}_\zeta^{(2)}\textbf{d}- 2\textbf{d}^{\mathsf{T}}\mbox{L}_\eta^{(2)}\boldsymbol{d}_\zeta+ 2\textbf{d}^{\mathsf{T}}\mbox{L}_\eta^{(2)} V  \mbox{L}_\zeta^{(2)}\textbf{d}\nonumber\\
&=&\frac{1}{2}\mbox{tr}(\partial_\zeta V  \mbox{L}_\eta^{(2)})+2\boldsymbol{d}_\eta^{\mathsf{T}} V ^{-1}\boldsymbol{d}_\zeta\nonumber\\
&=& \mathcal{F}_{\eta\zeta}~.
\end{eqnarray}

Similarly, for Eq.~\eqref{eq:j} we have
\begin{eqnarray}
\Im\left\{\mbox{tr}\left(\hat{\rho}_{\{\mu\}}\hat{\mathcal{L}}_\eta\hat{\mathcal{L}}_\zeta\right)\right\}&=& \textbf{d}^{\mathsf{T}}\mbox{L}_\eta^{(2)}\Omega\textbf{L}_\zeta^{(1)}+ 2\textbf{d}^{\mathsf{T}}\mbox{L}_\eta^{(2)}\Omega \mbox{L}_\zeta^{(2)}\textbf{d}+ \textbf{L}_\eta^{(1)\mathsf{T}}\Omega \mbox{L}_\zeta^{(2)}\textbf{d}+2\mbox{tr}\left(\Omega \mbox{L}_\zeta^{(2)} V  \mbox{L}_\eta^{(2)}\right)+ \frac{1}{2}\textbf{L}_\eta^{(1)\mathsf{T}}\Omega\textbf{L}_\zeta^{(1)}\nonumber\\
&=&2\textbf{d}^{\mathsf{T}}\mbox{L}_\eta^{(2)}\Omega V ^{-1}\boldsymbol{d}_\zeta- 2\textbf{d}^{\mathsf{T}}\mbox{L}_\eta^{(2)}\Omega V ^{-1}\boldsymbol{d}_\zeta+ 2\mbox{L}_\eta^{(2)}\Omega \mbox{L}_\zeta^{(2)}\textbf{d}+2\boldsymbol{d}_\eta^{\mathsf{T}} V ^{-1}\Omega \mbox{L}_\zeta^{(2)}\textbf{d}- 2\textbf{d}^{\mathsf{T}}\mbox{L}_\eta^{(2)}\Omega \mbox{L}_\zeta^{(2)}\textbf{d} \nonumber\\
&~&+2\mbox{tr}\left(\Omega L_\zeta^{(2)} V  \mbox{L}_\eta^{(2)}\right) +2\boldsymbol{d}_\eta^{\mathsf{T}} V ^{-1}\Omega V ^{-1}\boldsymbol{d}_\zeta -2\boldsymbol{d}_\eta^{\mathsf{T}} V ^{-1}\Omega \mbox{L}_\zeta^{(2)}\textbf{d}-
2\textbf{d}^{\mathsf{T}}\mbox{L}_\eta^{(2)}\Omega V ^{-1}\boldsymbol{d}_\zeta+2\textbf{d}^{\mathsf{T}}\mbox{L}_\eta^{(2)}\Omega \mbox{L}_\zeta^{(2)}\textbf{d}\nonumber\\
&=&2\mbox{tr}\left(\Omega \mbox{L}_\zeta^{(2)} V  \mbox{L}_\eta^{(2)}\right) +2\boldsymbol{d}_\eta^{\mathsf{T}} V ^{-1}\Omega V ^{-1}\boldsymbol{d}_\zeta \nonumber \\
&=&\mathcal{J}_{\eta\zeta}~.
\end{eqnarray}

In conclusion, to summarize, we have shown that
\begin{eqnarray}
\mbox{tr}\left(\hat{\rho}_{\{\mu\}}\hat{\mathcal{L}}_\eta\hat{\mathcal{L}}_\zeta\right)= \mathcal{F}_{\eta\zeta}+i\mathcal{J}_{\eta\zeta}~,
\end{eqnarray}
with
\begin{eqnarray}
\mathcal{F}_{\eta\zeta}&=&\frac{1}{2}\mbox{tr}(\partial_\zeta V  \mbox{L}_\eta^{(2)})+2\boldsymbol{d}_\eta^{\mathsf{T}} V ^{-1}\boldsymbol{d}_\zeta~,\\
\mathcal{J}_{\eta\zeta}&=&2\mbox{tr}\left(\Omega \mbox{L}_\zeta^{(2)} V  \mbox{L}_\eta^{(2)}\right) +2\boldsymbol{d}_\eta^{\mathsf{T}} V ^{-1}\Omega V ^{-1}\boldsymbol{d}_\zeta~.
\end{eqnarray}

This completes the proof of Theorem~\ref{thg}.

\section{Additional details on the estimation problem of Figure~1}\label{app:B}
For completeness, here we include the analytical expression of the QFI matrix for estimating the three parameters $\{\phi,x,y\}$ as described in Fig.~\ref{fig:Scheme}, using an input TMDSS $\hat{\rho}_0$ of the form given by Eq.~(\ref{eq:tmss}), with $r \geq 0$,
$\Re\left[\alpha\right] = \Re\left[\beta\right] = 0$, and $\Im\left[\alpha\right] = \Im\left[\beta\right]$.

The QFI matrix for the considered problem takes the form:
\begin{equation}\label{QFIsupp}
{\cal F}=\left(
\begin{array}{ccc}
 {\cal F}_{\phi \phi} & {\cal F}_{\phi x} & {\cal F}_{\phi y} \\
 {\cal F}_{\phi x} & {\cal F}_{x x} & {\cal F}_{x y} \\
{\cal F}_{\phi y} & {\cal F}_{x y} & {\cal F}_{y y} \\
\end{array}
\right)\,,
\end{equation}
where, using Eq.~(\ref{eq:gqfim}), we have:
\begin{eqnarray*}
 {\cal F}_{\phi \phi} &= & \frac{2 |\alpha|^2 x \left[x \sinh (2 r)+x \cosh (2 r)+y\right]}{2 x y \cosh (2 r)+x^2+y^2}\,,\\
 {\cal F}_{x x} & = & \frac{1}{2 x^2}\left\{\frac{\left(x^2-y^2+1\right)^2}{2 x y \cosh (2 r)+x^2+y^2-1}-\frac{\left[(x-y)^2+1\right] \left[(x+y)^2+1\right]}{2 x y \cosh (2 r)+x^2+y^2+1}+\frac{4 |\alpha|^2 x}{x+y}+2\right\}+\frac{2 |\alpha|^2 \left(e^{2 r}-1\right)}{(x+y) \left(e^{2 r} y+x\right)}\,,\\
 {\cal F}_{y y } & = & \frac{2 \left[x^2 \cosh (4 r)+2 x y \cosh (2 r)+y^2+1\right]}{2 x y \left[2 \cosh (2 r) \left(x^2+y^2\right)+x y \cosh (4 r)\right]+x^4+4 x^2 y^2+y^4-1}\,,\\
 {\cal F}_{\phi x} & = & 0\,, \\
 {\cal F}_{\phi y} & = & 0\,, \\
 {\cal F}_{x y} & = & \frac{2  \left(x^2+y^2+1\right) \cosh (2 r) +4 x y}{2 x y \left[2 \cosh (2 r) \left(x^2+y^2\right)+x y \cosh (4 r)\right]+x^4+4 x^2 y^2+y^4-1} \,.
\end{eqnarray*}

In the high input energy limit, $\bar n = \sinh^2(r) + |\alpha|^2 \gg 0$, the QFI matrix (\ref{QFIsupp}) can be approximated as
\begin{equation}\label{QFIsupphighn}
{\cal F} \approx \left(
\begin{array}{ccc}
 \frac{2px{\bar n}}{y}+c_\phi  & 0 & 0 \\
 0 & \frac{2p{\bar n}}{xy}+c_x  & O({\bar n}^{-1}) \\
0 & O({\bar n}^{-1}) & \frac{1}{y^2} \\
\end{array}
\right)\,,
\end{equation}
where $c_\phi$ and $c_x$ are some constants. From this we see that the nonzero off-diagonal term scales as ${\cal F}_{xy} \sim {\bar n}^{-1}$, thus vanishing in the limit ${\bar n} \gg 0$, in which case the compatibility condition (iii) is asymptotically fulfilled, as stated in Sec.~\ref{sec:example}. We also see that the variances on estimating $\phi$ and $x$ scale as the standard quantum limit, $\{{\cal F}_{\phi \phi}^{-1}, {\cal F}_{xx}^{-1}\} \sim {\bar n}^{-1}$, while the variance on estimating the added noise $y$ tends to a constant depending on the parameter itself, ${\cal F}_{yy}^{-1} \sim y^2$.

\subsection{Optimizing the input state for each parameter independently in individual estimation}\label{app:B2}

\begin{figure}[h]
	\centering
	\includegraphics[width=0.6\linewidth]{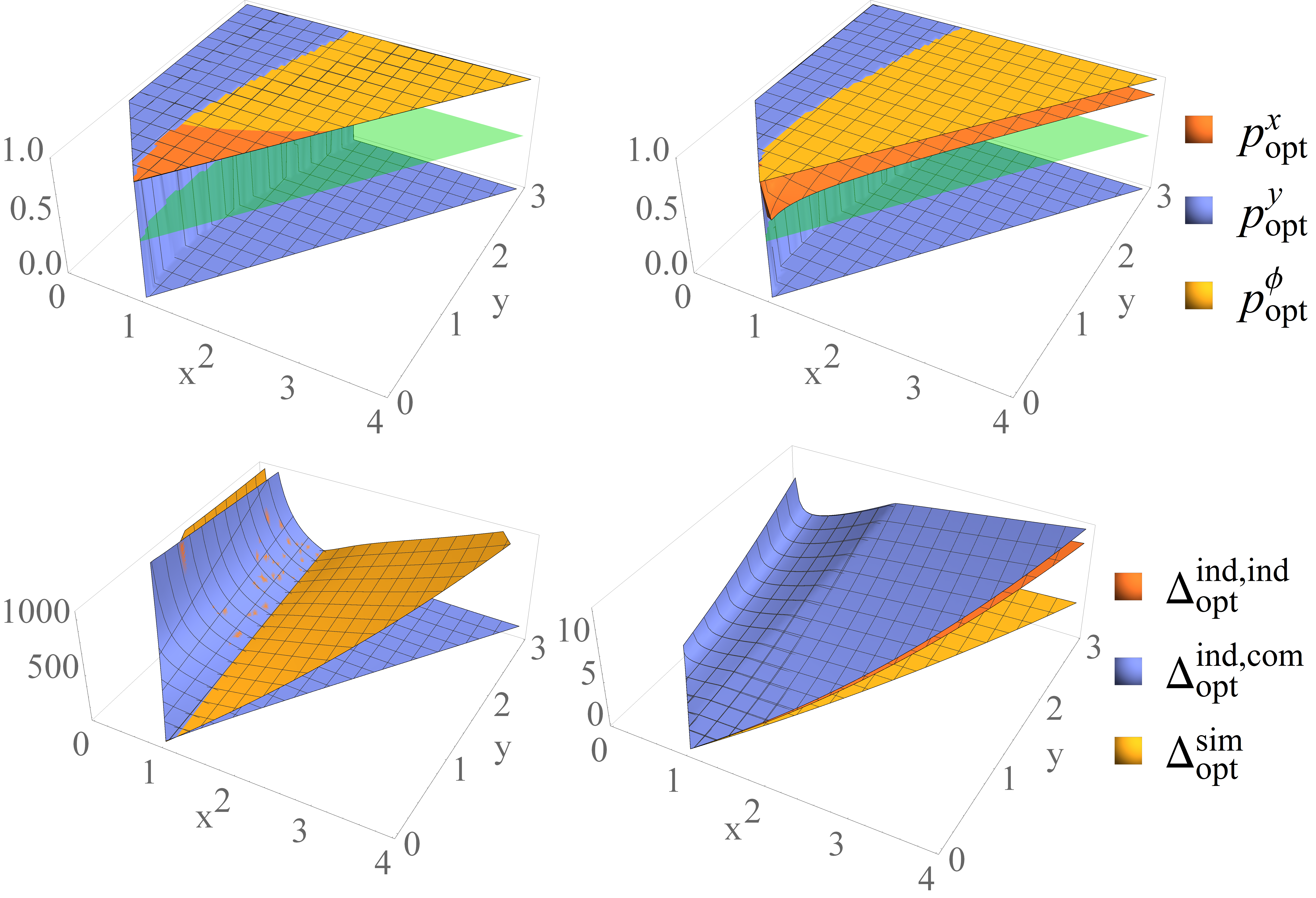}
	\caption{(Color online)
		Comparing optimizing the input state for each parameter independently (``ind,~\!ind''), to optimizing for the combined error (``ind,~\!com''), as in Fig.~\ref{fig:figures}.
		Left column: Analysis at  low energy, $\bar n = 0.005$. Right column: Analysis at  higher energy,  $\bar n = 5$.
		Top: Optimal proportion of input energy to put in the displacement, $p_{\mathrm{opt}}$, comparing individual estimations of each parameter. The unmeshed (green online) plane marks $p_{\mathrm{opt}} = 1/2$, above which more energy should be used for displacement than squeezing.
		Bottom: Minimal achievable error $\Delta_\mathrm{opt}$ for the introduced independent estimation, in comparison with the two strategies discussed in Sec.~\ref{sec:example}.
		All the presented results are independent of the value of the unknown phase $\phi$. All the quantities plotted are dimensionless.}
	\label{fig:figuresInd}
\end{figure}

The analysis of the metrological scheme in  Fig.~\ref{fig:Scheme} provided in Sec.~\ref{sec:example} compares the strategies of estimating each parameter individually and estimating them simultaneously. For simplicity of presentation, the analysis of individual estimation presented in Sec.~\ref{sec:example} in fact optimizes the minimal total combined (``com'') variance associated with the estimation of the three parameters over the family of input states i.e.,
\begin{equation} \label{eq:individual-combined-min-error}
\Delta_{\mathrm{opt}}^{\mathrm{ind}} \equiv  \Delta_{\mathrm{opt}}^{\mathrm{ind, com}}   = \min_{\hat{\rho}_0} \enskip \sum_{\eta\in \{\mu\}}{{\mathcal{F}^{-1}_{\eta\eta}}}
\end{equation}
Realistically, one may expect that in such an estimation procedure, as each parameter is estimated in an independent experiment, a different optimal input state could be determined for each parameter and used in each corresponding experiment. This would in principle lead to a smaller total variance, resulting from the sum of the minimal variances optimized independently (``ind'') for each parameter, thus altering our optimization to the following:
\begin{equation}\label{eq:individual-independent-min-error}
\Delta_{\mathrm{opt}}^{\mathrm{ind, ind}} = \quad   \sum_{\eta\in \{\mu\}}\Delta_{\mathrm{opt}}^{\eta} \quad = \quad \sum_{\eta\in \{\mu\}}{{\min_{\hat{\rho_0}} \enskip \mathcal{F}^{-1}_{\eta\eta}}}
\end{equation}
We present here the results of this independent optimization, finding that a slight improvement in the ensuing individual estimation strategy is obtained but the conceptual conclusions reached in Sec.~\ref{sec:example}, including the qualitative comparison with the simultaneous estimation strategy, remain unchanged.

As stated in Sec.~\ref{sec:example}, the compatibility condition (i) depends only on the displacement of the initial state and may be written as $\left| \alpha \right|^2 = \left| \beta \right|^2 $. In Sec.~\ref{sec:example}, it is also found that the minimum combined variance $\Delta_{\mathrm{opt}}^{\mathrm{ind, com}}$, as defined in Eq.~(\ref{eq:individual-combined-min-error}), is achieved with states that have $ \Re\left[\alpha\right] = \Re\left[\beta\right] = 0 $ and $ \Im\left[\alpha\right] = \Im\left[\beta\right]$. We find that these same conditions minimize each independent variance $\Delta^{\eta}$, so that also the quantity $\Delta_{\mathrm{opt}}^{\mathrm{ind,ind}}$ defined in Eq.~(\ref{eq:individual-independent-min-error}) is minimized and the compatibility condition remains obeyed.

The proportion of the energy $p_{\mathrm{opt}}$  to dedicate to the displacement of the initial state (as opposed to squeezing the state) for minimizing the combined error is discussed in Sec.~\ref{sec:example} and shown in Fig.~\ref{fig:figures} (Top). This quantity changes when considering the errors optimized for each parameter independently, as  shown in Fig.~\ref{fig:figuresInd} (Top). We find that to estimate $\phi$ one should always have all energy in the displacement. For estimating $x$, there exists a boundary in the parameter space, either side of which all energy should go to displacement or all energy should go to squeezing. As the total energy is increased, this boundary shifts as more of the parameter space favours squeezed probes over displaced probes. At low energy, to estimate $y$, all energy should be dedicated to displacement, while as the energy is increased the ratio $p_\mathrm{opt}$ varies but never drops below $1/2$, so more energy should always be dedicated to displacement than squeezing.

This shows a difference from minimizing the combined error. In Fig.~\ref{fig:figuresInd} (Bottom), we explore the effect this has on the total minimum error. We find that, as $\Delta_{\mathrm{opt}}^{\phi}$ dominates, $\Delta_{\mathrm{opt}}^{\mathrm{ind, ind}}$ is only slightly smaller than $\Delta_{\mathrm{opt}}^{\mathrm{ind, com}}$, therefore the refinement of the individual estimation scheme investigated here has {\it de facto} very little effect on the total error and the behaviour it displays. We therefore conclude that using $\Delta_{\mathrm{opt}}^{\mathrm{ind}} \equiv \Delta_{\mathrm{opt}}^{\mathrm{ind, com}}$ is adequate for discussing the qualities of the individual estimation scheme in the analyzed example, as is done in Sec.~\ref{sec:example}.
\end{widetext}


%

\end{document}